\documentclass[twocolumn,showpacs,preprintnumbers,amsmath,amssymb]{revtex4}

\usepackage{graphicx}
\usepackage{dcolumn}
\usepackage{bm}

\begin{document}

\title{Phonon dispersion and electron-phonon interaction for YBa$_2$Cu$_3$O$_7$ from first-principles calculations}
\author{K.-P. Bohnen$^{1}$, R. Heid$^{1}$, M. Krauss$^{2}$}
\affiliation{$^{1}$Forschungszentrum Karlsruhe, Institut f\"ur Festk\"orperphysik, P.O.B. 3640, D-76021 Karlsruhe, Germany\\
$^{2}$BASF IT Services GmbH, D-67056 Ludwigshafen, Germany}
%

%

\date{\today}

\begin{abstract}
We present a first principles investigation of the lattice
dynamics and electron-phonon coupling of the high-T$_c$
superconductor YBa$_2$Cu$_3$O$_7$ within the framework of density
functional perturbation theory using a mixed-basis pseudopotential
method. The calculated phonon dispersion curves are in excellent
agreement with Raman, infrared and neutron data. Calculation of
the Eliashberg function $\alpha^2$F leads to a small
electron-phonon coupling $\lambda$=0.27 in disagreement with
earlier approximate treatments. Our calculations strongly support
the view that conventional electron-phonon coupling is not an
important contribution to superconductivity in high-T$_c$
materials.
\end{abstract}

\pacs{74.24.Kc, 74.72.Bk, 78.30.-j}
\maketitle

Since the discovery of superconductivity in copper oxides in 1986
many different mechanisms have been suggested to explain the high
transition temperatures observed in these materials
\cite{Chubukov}. However, even after more than 15 years of
intensive research no consensus has been reached apart from the
fact that conventional electron-phonon coupling alone cannot
explain the observed transition
temperatures\cite{Cohen,Rodriguez,Andersen}. Even the much simpler
question concerning the magnitude of electron-phonon coupling has
not been answered conclusively due to the fact that calculations
in the past treated the lattice dynamics only in a very
approximate way. Recent progress in the ab initio calculation of
the lattice dynamics has made it possible to overcome these
limitations.

Modern bandstructure calculations together with ab-initio
determination of the phonon dispersion, electron-phonon coupling
and solution of the Eliashberg equations have made it possible to
calculate transition temperatures and study in detail the
electron-phonon coupling. This has recently been done very
successfully for the newly discovered superconductor
MgB$_2$\cite{Bohnen}. For high-T$_c$ superconductors (HTSC) in the
past however only a few attempts have been made to proceed along
the above mentioned road map to a deeper understanding of the
electron-phonon coupling
\cite{Cohen,Rodriguez,Andersen,Savarsov,Wang,Draxl}. The most
involved ones are those concerned with La$_2$CuO$_4$ and doped
CaCuO$_2$. For La$_2$CuO$_4$ the complete phonon dispersion has
been calculated however no electron-phonon coupling was
determined. Complications arise from the fact that the
calculations are carried out for the high-T$_c$ tetragonal phase
while at low T$_c$ experimentally an orthorhombic phase is
observed. More important however is the fact that for undoped
La$_2$CuO$_4$ an anitferromagnetic non-metallic ground state is
observed while LDA calculations find a metallic ground state which
questions the reliability of the calculations. The system
CaCuO$_2$, which is much simpler, also requires hole doping which
has been taken into account in an average way, however only after
properly doping the Fermi surface becomes similar to those of
other high-T$_c$ superconductors. Thus these results are also
questionable. To our knowledge so far all other ab-initio
treatments are restricted to the calculation of phonon modes for
the $\Gamma$-point only. In addition, even these calculations are
not complete but have been restricted to modes of special symmetry
classes only. Due to these limitations estimates of the coupling
strength are not very accurate, thus asking for a more complete
treatment of the electron-phonon coupling in the whole Brillouin
zone (BZ).

To clarify the role of electron-phonon coupling in the copper
oxides we have studied as a prototype YBa$_2$Cu$_3$O$_7$ along the
above mentioned lines. This system has been selected since it is
the best experimentally studied high-T$_c$ superconductor with
detailed information available about the Fermi surface and the
lattice dynamics and in addition does not have the complications
of doping and/or structural phase transitions. Furthermore,
bandstructure calculations based on the local-density
approximation for exchange and correlation (LDA) have been
successful in describing the Fermi surface and account properly
for the metallic ground state \cite{Kouba}. For YBa$_2$Cu$_3$O$_7$
we have carried out a systematic study of the lattice dynamics and
the electron-phonon coupling using the mixed-basis pseudopotential
method \cite{Meyer}. For all atoms we have used well-tested
potentials of BHS-type \cite{Bachelet}. As valence states we
included Y4s, Y4p, Ba5s, Ba5p and O2s which led to fairly deep
potentials, but they could be dealt with very efficiently due to
the mixed-basis formulation. The wave functions were constructed
from localized s,p,d functions at the yttrium and barium sites, s
and p functions at the oxygen sites and d-functions at the copper
sites supplemented by plane waves with an energy cut-off of 20 Ry.
Detailed tests were carried out to ensure convergence with respect
to the number of plane waves as well as with respect to k-point
sampling. For the structural optimization 576 k-points in the BZ
together with a Gaussian smearing of 0.2 eV have been used. The
calculation of the phonon dispersion is based on a recently
developed mixed basis perturbation approach \cite{Heid} which also
allowed for the calculation of the electron-phonon coupling
\cite{HeidPini}. Since these parts are very k-point sensitive up
to 5184 k-points in the BZ have been used here. All our
calculations were carried out using the local exchange-correlation
potential of Hedin and Lundqvist \cite{Hedin}.

Since YBa$_2$Cu$_3$O$_7$ crystallizes in the orthorhombic crystal
structure optimization requires the determination of 8 parameters
(the length of the axis a, b, c and 5 internal positions)(see Fig.
1). In table I our results are summarized and compared to all-
electron results \cite{Kouba} as well as experimental data
\cite{Kaldis,Schweiss}.

\begin{figure}
\centerline{\includegraphics[height=2.5in]{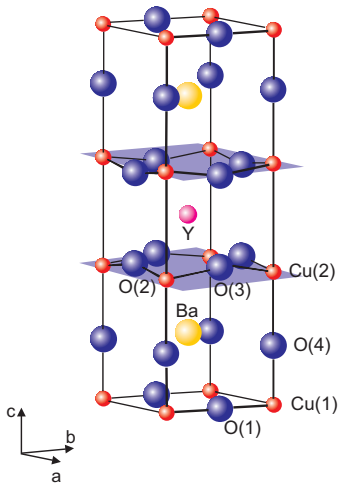}}
\caption{Crystal structure of YBa$_2$Cu$_3$O$_7$.}
\label{firstfigure}
\end{figure}

The two calculations give very similar results and the differences
are well within the limit of numerical accuracy. The calculated
equilibrium volume turns out to be 4\% to 5\% smaller than the
experimental one in agreement with previous calculations
\cite{Cohen,Kouba} and general trends observed for
LDA-calculations.

\begin{table}
\caption{Comparison of calculated and measured structural parameters of
YBa$_2$Cu$_3$O$_7$.}
\begin{ruledtabular}
\begin{tabular}
{cccc}
& this work & FLAPW\cite{Kouba} & Exp.\cite{Kaldis,Schweiss}\\
\hline Vol ({\AA}$^{3}$) & 165.6 & 163.1 & 172.32\\
c/a & 3.015 & 3.015 & 3.056\\
b/a & 1.021 & 1.012 & 1.018\\
Z$_{Ba}$ & 0.1819 & 0.1817 & 0.1837\\
Z$_{Cu(2)}$ & 0.3495 & 0.3507 & 0.3546\\
Z$_{O(2)}$ & 0.3770 & 0.3770 & 0.3783\\
Z$_{O(3)}$ & 0.3761 & 0.3765 & 0.3779\\
Z$_{O(4)}$ & 0.1627 & 0.1619 & 0.1592\\
\end{tabular}
\end{ruledtabular}
\end{table}

Having determined the equilibrium structure we now proceed to the
calculation of phonon modes. For comparison with other
calculations as well as experimental studies we concentrate first
on the $\Gamma$ point. Here, besides neutron scattering studies
also Raman and infrared data are available. Table II summarizes
the even symmetry modes and compares our results with those of an
all-electron study \cite{Draxl}(full potential linearized
augmented plane-wave calculations (FLAPW)) which is to our
knowledge the most extensive phonon study available up to now as
well as with experimental data \cite{Raman,Pini}. As pointed out
in connection with the FLAPW-study structure optimization has an
important effect on phonon frequencies. This has also been found
in our study. The agreement between both theoretical treatments
and the Raman data is very satisfactory. The pseudopotential
treatment is of equal quality as the FLAPW-method, in addition it
is computationally less demanding and allows thus the calculation
of odd modes as well which are given in table III together with
neutron- and IR-data \cite{Pini,Bernhard}.

\begin{table} \caption{Calculated and measured Raman modes. Besides the frequency (in meV) also
the dominant eigenvector character is given.}
\begin{ruledtabular}
\begin{tabular}{llccc}
& Mode & this work & FLAPW\cite{Draxl} & Exp.\cite{Raman}\\
\hline
A$_{1g}$ & Ba & 14.8 & 15.3 & 14.4-14.8\\
& Cu(2) & 18.8 & 18.2 & 18.0-18.6\\
& O(2)-O(3) & 42.1 & 41.9 & 41.5\\
& O(2)+O(3) & 50.4 & 52.3 & 53.9-54.6\\
& O(4) & 58.6 & 60.4 & 61.1-62.0\\
\hline
B$_{2g}$ & Ba & 8.0 & 8.1 & 8.7\\
& Cu(2) & 17.4 & 17.6 & 17.6\\
& O(4) & 27.8 & 27.5 & 26.0\\
& O(3) & 47.9 & 48.2 & 45.9\\
& O(2) & 71.4 & 73.5 & 71.8\\
\hline
B$_{3g}$ & Ba & 9.8 & 9.8 & 10.3\\
& Cu(2) & 17.4 & 17.5 & 17.4\\
& O(4) & 36.6 & 36.3 & 37.6\\
& O(2) & 45.3 & 46.1 & 46.9\\
& O(3) & 65.2 & 67.7 & 65.2\\
\end{tabular}
\end{ruledtabular}
\end{table}

\begin{table} \caption{Calculated and measured odd symmetry modes for the $\Gamma$ point in meV.}
\begin{ruledtabular}
\begin{tabular}
{llccc}
& Mode & Theory & Neutron & IR\\
\hline
B$_{1u}$ & Ba & 16.1 & - & -\\
& Cu(1) & 20.7 & 19.4 & 19.2\\
& Y/Cu(2) & 23.8 & 23.9 & 23.9\\
& O(2)-O(3) & 27.0 & - & silent\\
& O(1) & 29.8 & - & 34.1\\
& O(2)+O(3) & 40.3 & 37.2 & 37.2\\
& O(4) & 68.2 & 68.1 & 70.3\\
\hline
B$_{2u}$ & Y/Ba & 10.3 & 10.2 & -\\
& O(1) & 16.4 & 15.6 & -\\
& Cu(1) & 19.0 & 19.5 & -\\
& Y/Cu(2) & 21.8 & 24.0 & 23.5\\
& O(4) & 33.5 & 35.0 & 34.7\\
& O(3) & 44.6 & 44.0 & 44.9\\
& O(2) & 71.9 & 74.3 & 74.0\\
\hline
B$_{3u}$ & Y/Ba & 10.0 & 10.1 & -\\
& Cu(1) & 19.7 & 19.5 & -\\
& Y/Cu(2) & 21.2 & 23.4 & 23.4\\
& O(4) & 41.7 & - & 42.7\\
& O(2) & 42.6 & 44.5 & 44.9\\
& O(1) & 61.4 & 59.4 & 59.2\\
& O(3) & 65.6 & 68.0 & 67.6\\
\end{tabular}
\end{ruledtabular}
\end{table}

So far, we have discussed $\Gamma$-point modes only, however, a
reliable calculation of electron-phonon coupling requires the
knowledge of the complete phonon dispersion throughout the BZ.
Results for different directions in the BZ obtained using our
mixed basis perturbation approach \cite{Heid} are presented in
Figs.~2 and 3. The modes are classified according to symmetry. For
comparison also experimental data points are indicated. The
dispersion curves were obtained by determining the dynamical
matrix on a (4,4,2) reciprocal lattice grid in the BZ. Keeping in
mind the complexity of the system the results are in excellent
agreement with the available neutron data.

\begin{figure}
\centerline{\includegraphics[height=2.5in]{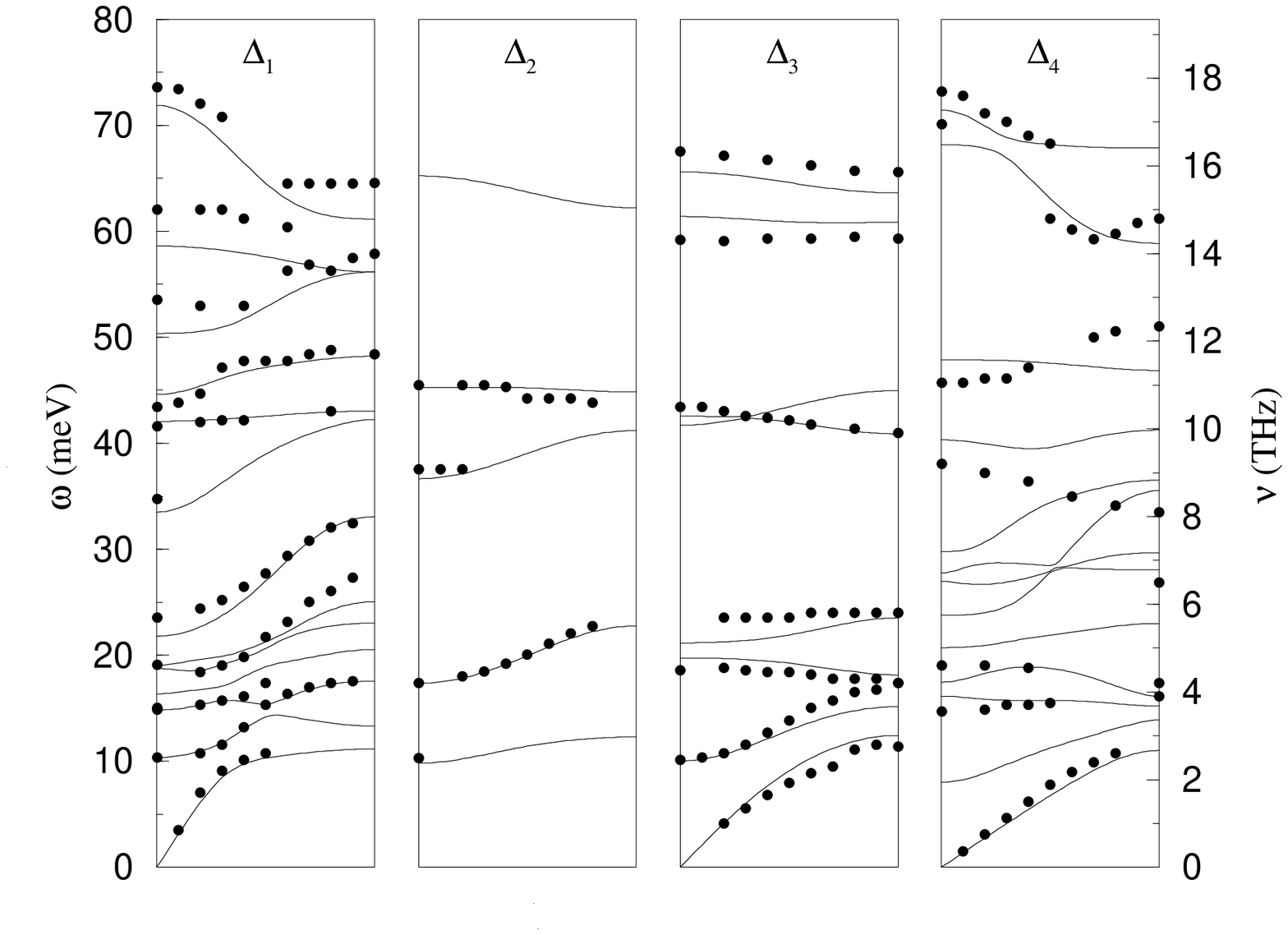}}
\caption{Phonon dispersion along [100] direction. The modes have
been classified according to symmetry. Lines represent theoretical
results while dots are neutron data \cite{Pini}.}
\label{secondfigure}
\end{figure}

\begin{figure}
\centerline{\includegraphics[height=2.5in]{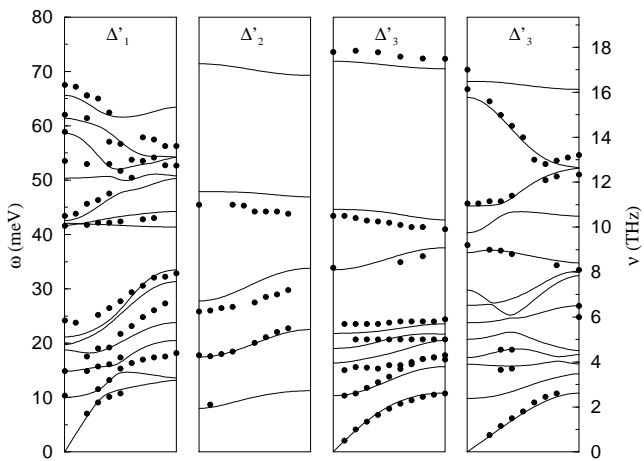}}
\caption{Phonon dispersion along [010] direction. The modes have
been classified according to symmetry. Lines represent theoretical
results while dots are neutron data \cite{Pini}.}
\label{thirdfigure}
\end{figure}

A number of points are worth mentioning in connection with the
dispersion curves. The strong renormalization in respect to
YBa$_2$Cu$_3$O$_6$ of the top-most mode of $\Delta_1$-symmetry
along [100] shows up quite naturally in these calculations while
model studies had extreme difficulties in reproducing this
behavior. The eigenvector shows that it is an oxygen
bondstretching mode in the Cu-O plane. For experimentalists one of
the major obstacles in the past has been the lack of untwinned
single crystals of YBa$_2$Cu$_3$O$_7$ which made sometimes the
assignment of a measured phonon to the (100) or (010)-direction
very difficult. Some of these ambiguous cases have been reassigned
in view of the calculations and in agreement with very recent
experimental studies \cite{Reznik}. Concerning the role of the
Cu-O chains we found that most of these modes are in the range
between 10 and 20 meV and only oxygen modes along the bond
direction are high-lying ones with frequencies around 60 meV. No
indication of a possible instability of the bond-bending chain
modes (as proposed by earlier calculations \cite{Cohen})was found
in agreement with experiment. Consistent with the expectation of
weak coupling perpendicular to the Cu-O-planes the dispersion in
(001) direction is very small.

In the upper panel of Fig. 4 we show the generalized phonon density-of-states which
should be compared with the neutron-scattering results presented
in Refs. \cite{Renker,Arai}. Peaks and even shoulder positions
agree quite well, as also the frequency cut off of 75 to 80 meV.
The ratio of peak intensities differs substantially between the
two measurements which is most likely due to the sensitivity of
intensities to the background corrections. This prohibits
comparison with our calculation.

\begin{figure}
\centerline{\includegraphics[height=2.8in]{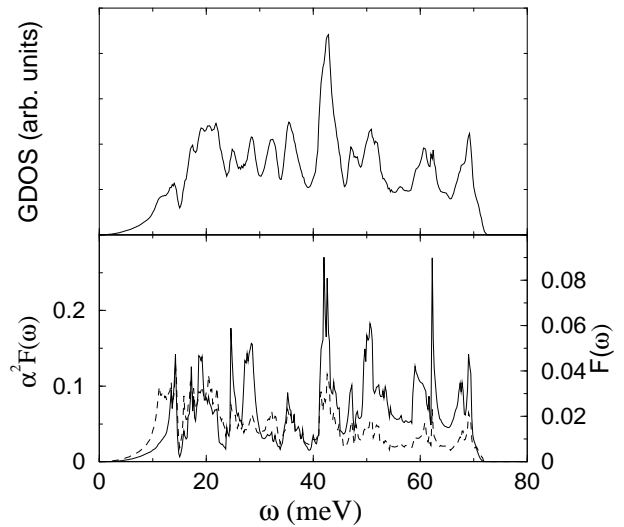}}
%
\caption{Upper panel: Generalized phonon density of states. A broadening of 0.5
meV has been used.
Lower panel: Eliashberg function $\alpha^2$F($\omega$) (full line) and
phonon density of states F($\omega$)(dotted line).} \label{fourthfigure}
\end{figure}

To address the superconducting properties we have calculated the
so-called Eliashberg function $\alpha^{2}$F($\omega$)
\cite{Eliashberg}, using a very fine (24,24,4) k-point grid in
performing the Fermi-surface average of the electron-phonon matrix
elements \cite{Bauer}. Extensive tests have shown that this grid
is sufficient for our calculations. Results are presented in Fig.
4 (lower panel). Comparing the phonon density of states F($\omega$) and
$\alpha^{2}$F($\omega$) we find a substantial shift of weight for
$\alpha^{2}$F($\omega$) to higher frequencies.

From the Eliashberg function we calculate the electron phonon
coupling constant
$
\lambda = 2 \int _{0}^{\infty} d\omega \alpha^{2}F(\omega) / \omega
$
which gives $\lambda$=0.27. For the logarithmically averaged
phonon frequencies as defined in Ref. \cite{Allen} we find 27.7
meV. After solving the isotropic gap equation for $\mu$*=0 we
obtain only a T$_c$ of $\sim$ 2 K. Any realistic $\mu$* suppresses
superconductivity completely.

The very small value of $\lambda$ is in substantial disagreement
with earlier estimates based on $\Gamma$-point results only.
Analyzing the behavior of $\lambda$ throughout the BZ we find it
to be varying by a factor of two only, however, variation as
function of individual modes for given q is much larger. For the
$\Gamma$-point for example modes of A$_g$ and B$_{1u}$-symmetry
account for 80\% of $\lambda$. The largest contributions come from
modes in the copper-oxygen planes as expected. Previous estimates
of $\lambda$ assumed that all modes at $\Gamma$ contribute with
similar strength as the A$_g$-modes. This elucidates why
$\lambda$-values of the order of 1 or even larger have been
obtained. However, this is a pure artefact of the unjustified
assumption.
%
%

Raman measurements at $\Gamma$ allowed for the determination of
the anomalous frequency shifts of lattice vibrations occurring
when cooling a superconductor below T$_c$ \cite{Thomsen}. Based on
a theory by Zeyher and Zwicknagl for the phonon-self energy it
could be shown that values of $\lambda_1$=0.02 for the A$_g$ mode
around 42 meV and $\lambda_2$=0.01 for the A$_g$-mode around 54
meV, as determined by an LMTO-calculation \cite{Rodriguez}, are
consistent with the observed experimental frequency shifts. Our
calculation give slightly larger values of $\lambda_1\approx$0.06
and $\lambda_2\approx$0.02 which are equally consistent with the
measurements due to the approximations inherent in the calculation
of the phonon-self energy \cite{Zeyher}.

So far we have discussed only s-wave pairing, however many
experimental facts support d-wave pairing \cite{Wollmann}. In the
much simpler system CaCuO$_2$ the influence of s- versus d-pairing
has been studied \cite{Savarsov} and it has been found that both
channels give very similar results for $\lambda$ which should also
hold for YBa$_2$Cu$_3$O$_7$. One other question which has not been
discussed so far is the influence of correlations beyond LDA. Here
the strongest argument in favor of our calculations is the fact
that the calculated phonon dispersion is in very good agreement
with experiment. If the electron-phonon coupling would be very
strongly dependent on correlations this should show up in
significant discrepancies between calculated and measured phonon
modes. The good agreement supports the claim that correlation
effects are of minor importance for electron-phonon coupling.

In summary, we have presented here ab-initio results for the
lattice dynamics of YBa$_2$Cu$_3$O$_7$ which are in excellent
agreement with available Raman, infrared and neutron data. Based
on the knowledge of the complete phonon spectrum the
electron-phonon coupling has been calculated. The electron-phonon
coupling constant turns out to be very small. These calculations
strongly support the view that in high-T$_c$ materials
conventional electron-phonon coupling is not an important
contribution to superconductivity.

The authors thank 
Dr. L. Pintschovius and Dr. W.
Reichardt for numerous discussions.



\end{document}